\documentclass[aps,prd,showpacs,twocolumn]{revtex4}
\usepackage{amssymb}
\usepackage{amsmath}
\usepackage{tikz}
\usepackage{color}
\usepackage{esdiff}
\usetikzlibrary{calc}
\usetikzlibrary{arrows}

\begin{document}
\title{An alternative formulation of Classical Mechanics based on an analogy with Thermodynamics}
\author{Gin\'{e}s R. P\'{e}rez Teruel$^1$}\email{gines.landau@gmail.com}
\affiliation{$^1$Departamento de F\'{i}sica Te\'{o}rica, Universidad de Valencia, Burjassot-46100, Valencia, Spain}
\email{gines.landau@gmail.com}

\begin{abstract}
We study new Legendre transforms in classical mechanics and investigate some of their general properties. The behaviour of the new functions is analyzed under coordinate transformations.When invariance under different kinds of transformations are considered, the new formulation is found to be completly equivalent to the usual Lagrangian formulation, recovering well established results like the conservation of the angular momentum. Furthermore, a natural generalization of the Poisson
Bracket is found to be inherent to the formalism introduced. On the other hand, we find that with a convenient redefinition of the Lagrangian, $\mathcal{L}^{\prime}=-\mathcal{L}$, it is possible to establish an exact one-to-one mathematical correspondence between the thermodynamic potentials and the new potentials of classical mechanics
\end{abstract}
\maketitle
\section{Introduction}
\label{sec:intro}
The Legendre transform is a powerful tool with important applications in virtually every branch of physics. It is not the purpose of this letter to study the nature of the Legendre transform itself. For this the reader is referred to other works \cite{Tra,Zia,Adl,Tou}, which treat various aspects and subleties of the Legendre transform in all mathematical rigor. In this paper we will deal with new functions in classical mechanics, which we derived by Legendre transforms from the usual Lagranian and Hamiltonian formulations, employing only the mathematical machinery that is customary in these theories.

As is well known, the Hamiltonian ($H$) can be derived from a regular Lagrangian ($\mathcal{L}$) via a Legendre transform, namely
\begin{equation}
  \label{eq:hamiltonian}
  H(q_i,p_i)=\sum_{i=1}^{N} \dot q_i p_i - \mathcal{L}(q_i, \dot q_i)\,
\end{equation}
It is typically assumed that this transformation is the only Legendre transform that is useful in classical mechanics. However, in thermodynamics we have {\it four} fundamental functions, relating the {\it four} natural variables $T$, $P$, $V$, $S$. Each of these functions can be obtained from any of the other three by a Legendre transformation. One naturally wonders what applications the {\em unused} ``mechanical potentials" might have. The subject of the following letter is to study the extension of classical mechanics to one with two additonal dynamical potentials that parallel the Gibbs and Helmholtz free energies in thermodynamics, thus completing the couple formed by $H$ and $\mathcal{L}$.

With the introduction of two new functions, we will have a set of four ``mechanical potentials'' obtained by transformations of the ``four '' dynamical variables $p_i$, $q_i$, $\dot q_i$, $\dot p_i$ in exactly the same fashion that the four thermodynamic potentials exchange their dependence upon the natural variables by Legendre transforms.  
The rest of the paper is organized as follows. In sec. \ref{sec:gen-Leg-Transform} we introduce {\it new} Legendre transforms which involve the time derivative of the canonical momentum $\dot p_i$. In sec. \ref{sec:noether-theorem} we analyze these Legendre transforms and derive some results in connection with them. Finally in sec. \ref{sec:one-to-one} we study the formal mathematical analogies between thermodynamics and classical mechanics.
\section{Beyond the standard Legendre Transform}
\label{sec:gen-Leg-Transform}
Here by the {\it usual} Legendre transform we refer to a transformation that through a regular Lagrangian leads to the Hamiltonian of a classical system i.e. transformations of the type given by (\ref{eq:hamiltonian}). Thus, our aim is to construct functions beyond $H(q_i,p_i)$ and $\mathcal{L}(q_i,\dot q_i)$ that not only depend on $q_i$, $\dot q_i$, and $p_i$, but also depend upon $\dot p_i$. Thus, let us consider the pair of functions
\begin{equation}
  \label{eq:functions}
  J(\dot q_i,\dot p_i)\quad\mbox{and}\quad Q(p_i,\dot p_i)
\end{equation}
where
\begin{equation}
  \label{eq:generalized-force}
  \dot p_i=-\frac{\partial H}{\partial q_i}
\end{equation}
is the time derivative of the canonical momentum.
For reasons that will be clarified later, we consider it convenient to take the Hamiltonian as the starting point of our analysis, in contrast to \cite{Per}, where we introduced and defined these functions directly from the Lagrangian. Of course, both treatments are compatible and consistent, but it turns out to be more suitable to start from the Hamiltonian to make contact with the thermodynamic structures, as we will see in the following sections.

In order to guarantee that the functions introduced before are well defined Legendre transforms, the Hamiltonian $H(q_i,p_i)$, must be regular not only in the generalized coordinates but in the generalized momentum as well. Once these conditions are satisfied the following transformations can be defined
\begin{eqnarray}
  \label{eq:transformations1}
  J(\dot q_i,\dot p_i) & = & \sum_{i=1}^{N} (\dot p_i q_i-\dot q_i p_i )+H(q_i,p_i)\,\\
  \label{eq:transformations2}
  Q(p_i,\dot p_i) & = & \sum_{i=1}^{N} \dot p_i q_i + H(q_i,p_i)\,
\end{eqnarray}
From these definitions it follows that
\begin{eqnarray}
dJ=\sum_{i=1}^{N}\left(q_i d\dot p_i-p_i d\dot q_i\right)\,\nonumber\\
dQ=\sum_{i=1}^{N}\left(q_i d\dot p_i+\dot q_i dp_i\right)\,
\end{eqnarray}
which implies
\begin{align}
  \label{eq:constraints1}
  \frac{\partial J}{\partial \dot q_i}&=-p_i
  \qquad
  \frac{\partial J}{\partial \dot p_i}=q_i \\
  \label{eq:constraints2}
  \frac{\partial Q}{\partial \dot p_i}&=q_i
  \qquad
  \frac{\partial Q}{\partial p_i}=\dot q_i
\end{align}
From eqs. (\ref{eq:constraints2}) the following identities, involving the function $Q(p_i,\dot p_i)$, can be written
\begin{equation}
  \label{eq:euler-lag-likeEQS}
  \frac{\partial Q}{\partial p_i}-\frac{d}{dt}\frac{\partial Q}
  {\partial \dot p_i}=0\,
\end{equation}
It is interesting to note that they have the same form that Euler-Lagrange equations and will be shown to be useful in the next section.

\section{Noether's Theorem and symmetry properties of $\pmb{J}$ and $\pmb{Q}$}
\label{sec:noether-theorem}
Noether's theorem states that to any symmetry of the Lagragian corresponds a conserved quantity. In particular, when the Lagrangian does not have an explicit time dependence the usual Legendre transform, the Hamiltonian $H(q_i,p_i)$, is constant. We will show in detail that this statement holds not only for $\mathcal{L}(q_i(t),\dot q_i(t))$ but also for the functions $J(\dot q_i(t),\dot p_i(t))$ and $Q(p_i(t),\dot p_i(t))$ that were introduced in the previous section.

If we evaluate the time derivative of the $Q$ function
\begin{align}
  \label{eq:Q-variation-long1}			
  \frac{dQ}{dt}&=\sum_{i=1}^{N} \left(\frac{\partial Q}{\partial \dot p_i}\ddot p_i+\frac{\partial Q}{\partial p_i}\dot p_i\right)\nonumber\\
  &=\frac{d}{dt}\left(\sum_{i=1}^{N}\frac{\partial Q}{\partial \dot p_i} \dot p_i\right)
  + \sum_{i=1}^{N}\dot p_i\left(\frac{\partial Q}{\partial p_i}
  - \frac{d}{dt}\frac{\partial Q}{\partial \dot p_i}\right)\,
\end{align}
The last term vanishes due to (9) and finally we obtain the following result
\begin{align}  
  \frac{d}{dt}\left(Q-\sum_{i=1}^{N}\frac{\partial Q}{\partial \dot p_i} \dot p_i\right)=\frac{dH}{dt}=0\,
\end{align}
where we have used that, $\displaystyle Q-\sum_i\frac{\partial Q}{\partial \dot p_i}\dot p_i=H$, in accordance with equations (5) and (8).
It is worth noting that an identical constraint can be derived making use of the other dynamical function $J(\dot q_i,\dot p_i)$. After a bit of algebra
\begin{align}
  \frac{dJ}{dt}&=\sum_{i=1}^{N}\left(\frac{\partial J}{\partial\dot q_i}\ddot q_i+\frac{\partial J}{\partial \dot p_i}\ddot p_i\right)=\sum_{i=1}^{N}\left(q_i\ddot p_i-p_i\ddot q_i\right)\nonumber\\
	&=\frac{d}{dt}\left(\sum_{i=1}^{N}q_i\dot p_i- p_i\dot q_i\right)\,
\end{align}
This result automatically implies that
\begin{align}
\frac{	d}{dt}\left(J-\sum_{i=1}^{N}(q_i\dot p_i-p_i\dot q_i)\right)=0\,
\end{align}
Taking into account the definition of $J(\dot
q_i,\dot p_i)$ given in (4) the last expression is nothing but the conservation of the Hamiltonian once again.
\subsection{ Some examples: Invariance under rotations and conservation of  the Angular Momentum.The harmonic oscillator}
Now we are interested in the existence of conserved quantities linked with symmetries of these dynamical functions $J(\dot q_i,\dot p_i), Q(p_i,\dot p_i)$ that we have introduced, but from a more general point of view. For instance, let us consider an arbitrary transformation of the $Q(p_i,\dot p_i)$ function. The variation $\delta Q$ will be equal to
\begin{equation}
  \label{eq:Q-variation}
  \delta Q =\sum_{i=1}^{N}\left(\frac{\partial Q}{\partial \dot p_i}\delta \dot p_i +
  \frac{\partial Q}{\partial p_i}\delta p_i\right)\,
\end{equation}
This can be rewritten according to
\begin{align}
  \label{eq:Q-variation-long2}
  \delta Q &=
  \frac{d}{dt}\left(
    \sum_{i=1}^{N}\frac{\partial Q}{\partial \dot p_i}\delta p_i
  \right) -
  \sum_{i=1}^{N}\left(
    \frac{d}{dt}
    \frac{\partial Q}{\partial \dot p_i}
  \right)\delta p_i
  + \sum_{i=1}^{N}\frac{\partial Q}{\partial p_i}\delta p_i
  \nonumber\\
  &=\frac{d}{dt}\left(\sum_{i=1}^{N}\frac{\partial Q}{\partial \dot p_i}\delta p_i\right)
  + \sum_{i=1}^{N}\left(\frac{\partial Q}{\partial p_i}
  - \frac{d}{dt}\frac{\partial Q}{\partial \dot p_i}\right)\delta p_i\,
\end{align}
Due to (\ref{eq:euler-lag-likeEQS}) the second term vanishes, and (\ref{eq:Q-variation-long2}) gives the result
\begin{equation}
  \label{eq:deltaQ-final}
  \delta Q=\frac{d}{dt}
  \left(
    \sum_{i=1}^{N}\frac{\partial Q}{\partial \dot p_i}
    \delta p_i
  \right)\,
\end{equation}

Thus, if the transformation is a symmetry of the function $Q$ ($\delta
Q=0$), the quantity $\displaystyle R\equiv\sum_i(\partial Q/\partial \dot p_i) \delta p_i=\sum_iq_i\delta p_i$ is a conserved constant of motion.
It is worth noting that $\sum_iq_i\delta p_i$ will have dimensions of angular momenta if the canonical momentum is equal to the linear momentum. Indeed, in order to better understand the last result, let us consider an arbitrary rotation in three dimensions. Under a rotation of angle $\theta$ around the $z$ axis, the components of the momentum transform according to
\begin{equation*}
\begin{pmatrix}
p^{\prime}_{x}(\theta)\\
p^{\prime}_{y}(\theta)\\
p^{\prime}_{z}(\theta)
\end{pmatrix}
=
\begin{pmatrix}
cos\theta & sin\theta & 0 \\
-sin\theta & cos\theta& 0 \\
0 & 0 & 1
\end{pmatrix}
\begin{pmatrix}
p_x \\
p_y\\
p_z
\end{pmatrix}
\end{equation*}
\begin{eqnarray}
  \label{eq:transformations1}
  p^{\prime}_{x}(\theta)&=p_xcos\theta+p_ysin\theta\nonumber\\
  \label{eq:transformations2}
  p^{\prime}_{y}(\theta)&=p_ycos\theta-p_xsin\theta\
\end{eqnarray}

For an infinitesimal transformation
\begin{eqnarray}
  \delta p_x=p_y\nonumber\\
  \delta p_y=-p_x \
\end{eqnarray}
Therefore if $Q$ remains invariant under the infinitesimal rotation, we can derive directly from (16) the conservation of the angular momentum
\begin{align}
\frac{d}{dt}\left(\sum_{i=1}^{N}\frac{\partial Q}{\partial \dot p_i}\delta p_i\right)&=\frac{d}{dt}\left(\sum_{i=1}^{3}q_i\delta p_i\right)=\frac{d}{dt}\left(x\delta p_x+y\delta p_y\right)\nonumber\\
&=\frac{d}{dt}\left(xp_y-yp_x\right)=\frac{d}{dt}(L_{z})=0\ \,
\end{align}

Thus, we have been able to derive results that are identical to those obtained in the Lagrangian picture of classical mechanics. We can establish a one-to-one coorespondence to illustrate the main differences between the formulations.
\begin{table}[ht]
\caption{Behaviour under coordinate transformations} 
\centering				
\begin{tabular}{c c c c}		
\hline\hline				
Function & Variables & Transformation & Conserved quantities \\ [0.8ex] 
\hline 					
\\ [1ex]				
$\mathcal{L}$		
&					
$q_i$, $\dot q_i$
&					
$q_i\rightarrow q_i^{\prime}$
&					
$\displaystyle\sum_i\frac{\partial \mathcal{L}}{\partial \dot q_i}\delta q_i$ \\				
$Q$
&					
$p_i$, $\dot p_i$
&					
$\displaystyle p_i\rightarrow p_i^{\prime}$
&					
$\displaystyle\sum_i\frac{\partial Q}{\partial \dot p_i}\delta p_i$ \\
$H$
&					
$q_i, p_i$
&					
&  \\					
\\ [0.8ex]				
$J$
&					
$\dot q_i, \dot p_i$
&					
E.T.I.
&					
$\displaystyle J-\sum_i(q_i\dot p_i-p_i\dot q_i)$ \\
\\ [1ex] 				
\hline					
\end{tabular}
\label{table:nonlin}			
\end{table}

In the previous table, E.T.I denotes explicit time-independence. It is very interesting to note the similarities between $\mathcal{L}$ and $Q$. We can say that $Q$ is a Lagrangian-type function, because it satisfies a particular version of Euler-Lagrange equations (9)

\begin{eqnarray}
\displaystyle\frac{\partial \mathcal{L}}{\partial q_i}-\frac{d}{dt}\frac{\partial \mathcal{L}}{\partial \dot q_i}=0 \nonumber\\
\frac{\partial Q}{\partial p_i}-\frac{d}{dt}\frac{\partial Q}
  {\partial \dot p_i}=0\ \,
\end{eqnarray}
In addition, it has very similiar invariance properties under coordinate transformations
\begin{eqnarray}
  \delta\mathcal{L}=0 \Rightarrow \sum_{i=1}^{N}\frac{\partial \mathcal{L}}{\partial \dot q_i}\delta q_i=C\nonumber\\
  \delta Q=0 \Rightarrow \sum_{i=1}^{N}\frac{\partial Q}{\partial \dot p_i}\delta p_i=C^{\prime}\ \,
\end{eqnarray}

The reason of these mathematical similarities lies in the fact that the functions $\mathcal{L}$ and $Q$ differ only by a total derivative. Indeed, using (1) and (5) it is easy to see that

\begin{align}
\mathcal{L}(q_i,\dot q_i)&=\sum_{i=1}^{N} \dot q_i p_i -H=\sum_{i=1}^{N} \dot q_i p_i-\left(Q(p_i,\dot p_i)-\sum_{i=1}^{N} \dot p_i q_i \right)\nonumber\\
&=\frac{d}{dt}\left(\sum_{i=1}^{N} q_i p_i\right)-Q(p_i,\dot p_i)\,
\end{align}

From our point of view, the choice of one formulation or the another is only a matter of convenience, because they are dynamically equivalent and lead to the same physical results as we have proved. This is consistent with the known fact \cite{Gol} that assures the independence of classical mechanics under canonical transformations, which are more general than Legendre transforms.\

Having reached this point, we want to apply the theoretical framework developed with the introduction of the mechanical potentials $Q(p_i,\dot p_i)$ and $J(\dot q_i,\dot p_i)$ to some practical problem.Let us briefly discuss how they can be succesfully applied to the well known, one-dimensional harmonic oscillator. The Hamiltonian is
\begin{equation}
H(x,p)=\frac{p^2}{2m}+\frac{1}{2}kx^2 \,
\end{equation}\\
which provides the well known identities $\partial H/\partial p=\dot x$, and $\partial H/\partial x=kx=-\dot p=-f $. Now, in order to go from $H(x,p)$ to $Q(p,\dot p)$ we should take into account the definition given in eq. (5)
\begin{align}
Q(p,f)&=H(x,p)+fx=\frac{p^2}{2m}-\frac{1}{2}fx+fx=\frac{p^2}{2m}+\frac{1}{2}fx\nonumber\\
&=\frac{p^2}{2m}-\frac{f^2}{2k}\ \,
\end{align}

At this point, we need to make an important clarification. In this particular case, we have a conservative force that comes directly from a potential, and thus $Q(p,\dot p)\equiv Q(p,f)$ but in more general situations, where the potential has a dependence on the velocities, this is no longer true. Thus, in this particular case of the harmonic oscillator, we can replace $\dot p$ by $f$, and the equations of motion in the $Q(p,\dot p)$ picture can be obtained by application of the Euler-Lagrange type equations (9), particularized to this problem
\begin{equation}
 \frac{\partial Q}{\partial p}-\frac{d}{dt}\frac{\partial Q}
  {\partial f}=0\
\end{equation}
This yields 
\begin{equation}
p=-\frac{m}{k}\frac{df}{dt} \
\end{equation}
which looks a bit odd, but if we take the derivative of both sides of the last equation we get
\begin{equation}
\frac{d^2f}{dt^2}+\frac{k}{m}f=0 \
\end{equation}
This is exactly the second order differential equation that is expected for an harmonic oscillator of frequency given by $\omega^2=k/m$.
\subsection{Generalized Poisson Brackets}

If it exists an important concept in classical mechanics this is the Poisson Bracket, a binary operation invariant under canonical transformations that governs the time- evolution of Hamiltonian mechanics. This concept is defined as follows: given two functions in the phase space, $f(q_i,p_i)$, $g(q_i,p_i)$, the Poisson Bracket of  $f$, $g$, acquires the form

\begin{equation}
\{f,g\} = \sum_{i=1}^{N} \left( 
\frac{\partial f}{\partial q_{i}} \frac{\partial g}{\partial p_{i}} - \frac{\partial f}{\partial p_{i}} \frac{\partial g}{\partial q_{i}}\right)
\end{equation}

This binary operation is antisymmetric: $\{f,g\}=-\{g,f\}$ and it obeys the Jacobi identity: $\{f,\{g,h\}\} + \{h,\{f,g\}\} + \{g,\{h,f\}\} = 0$. Therefore, it is defining a Lie algebra, known as Poisson algebra.

It is worth noting that the Legendre transforms that we introduced in equations (4) and (5) involve a binary operation that can be viewed as a generalization of the Poisson Bracket. Indeed, taking into account the partial derivative identities (7), we can write the equation (4) as follows

\begin{align}
  \label{eq:transformations1}
  J(\dot q_i,\dot p_i)& = \sum_{i=1}^{N} (\dot p_i q_i-\dot q_i p_i )+H(q_i,p_i)\nonumber\\
	&=\sum_{i=1}^{N} \left( 
\frac{\partial J}{\partial \dot q_{i}} \frac{\partial H}{\partial p_{i}} - \frac{\partial J}{\partial \dot p_{i}} \frac{\partial H}{\partial q_{i}}\right)+H(q_i,p_i)\,
\end{align}

In the same fashion, we can find a relation among the two other mechanical potentials $\mathcal{L}$, $Q$ in terms of another binary operation. Combining equations (22) and (8) we obtain

\begin{align}
\mathcal{L}(q_i,\dot q_i)&=\frac{d}{dt}\left(\sum_{i=1}^{N} q_i p_i\right)-Q(p_i,\dot p_i)\nonumber\\
&=\sum_{i=1}^{N}\left( \dot q_i p_i+q_i\dot p_i\right)-Q(p_i,\dot p_i)\nonumber\\
&=-\sum_{i=1}^{N}\left( \frac{\partial \mathcal{L}}{\partial q_{i}} \frac{\partial Q}{\partial \dot p_{i}} - \frac{\partial \mathcal{L}}{\partial \dot q_{i}} \frac{\partial Q}{\partial p_{i}}\right)-Q(p_i,\dot p_i)\,
\end{align}

The binary operations that appear in these latter equations are very interesting objects, and in some sense can be interpreted as a natural generalization of the usual Poisson Bracket. The standard Poisson Bracket operates over functions defined in the same space, the phase space of coordinates $q_i, p_i$. Nevertheless, the binary operators in equations (29) and (30) are connecting functions that live in different spaces. In general, if we have two functions $f(x_i,y_i)$, $g(z_i, w_i)$ that take values in different spaces characterized by the coordinates: $x_i, y_i$ and $z_i,w_i$, it will be possible to define a Generalized Poisson Bracket that mixes the functions and their variables

\begin{equation}
\{f(x_i,y_i),g(z_i,w_i)\}^{\dagger}=\sum_{i=1}^{N}\left( \frac{\partial f}{\partial x_{i}} \frac{\partial g}{\partial  w_{i}} - \frac{\partial f}{\partial y_{i}} \frac{\partial g}{\partial z_{i}}\right)\
\end{equation}

It is straightforward to verify that the usual Poisson Bracket (28), is only a particular case of (31). This identification occurs when the functions are defined in the same space of configurations, the phase space. 
The reader may verify that if $x_i= z_i\equiv q_i$, and $y_i= w_i\equiv p_i$, then: $\{f,g\}^{\dagger}=\{f,g\}$

In summary, the compact expressions for (29) and (30) employing the Generalized Poisson Brackets will be

\begin{eqnarray}
J(\dot q_i,\dot p_i)=\{J,H\}^{\dagger}+H(q_i, p_i)\nonumber\\
\mathcal{L}(q_i,\dot q_i)=-\{\mathcal{L},Q\}^{\dagger}-Q(p_i,\dot p_i)\,
\end{eqnarray}\\

Where

\begin{equation}
\{J(\dot q_i,\dot p_i),H(q_i,p_i)\}^{\dagger}=\sum_{i=1}^{N} \left( 
\frac{\partial J}{\partial \dot q_{i}} \frac{\partial H}{\partial p_{i}} - \frac{\partial J}{\partial \dot p_{i}} \frac{\partial H}{\partial q_{i}}\right)
\end{equation}
\begin{equation}
\{\mathcal{L}(q_i,\dot q_i),Q(p_i,\dot p_i)\}^{\dagger}=\sum_{i=1}^{N}\left( \frac{\partial \mathcal{L}}{\partial q_{i}} \frac{\partial Q}{\partial \dot p_{i}} - \frac{\partial \mathcal{L}}{\partial \dot q_{i}} \frac{\partial Q}{\partial p_{i}}\right)
\end{equation}

\newpage
\subsection{The Variational Principle. Physical interpretation of the introduced functions}

As is well known, the Lagrangian admits a natural interpretation provided by Hamilton's principle, which is enough to derive the Euler-Lagrange equations under completely general assumptions.One naturally wonders if the other mechanical potentials could be employed in the same fashion, within the context of some sort of variational principle. Regarding the function $Q(p_i,\dot p_i)$, we will show that it is possible.

For this purpose, it is convenient to begin with the usual definition of the action integral in terms of the Lagrangian

\begin{equation}
\mathcal{S} = \int_{t_1}^{t_2} \mathcal{L}(q_i,\dot q_i) \, dt\,
\end{equation}

Now making use of eq. (22) we can express the action integral in terms of $Q(p_i,\dot p_i)$ and a total derivative

\begin{align}
\mathcal{S}& = \int_{t_1}^{t_2} \mathcal{L}(q_i,\dot q_i) \, dt = \int_{t_1}^{t_2} \left(\frac{d}{dt}\left(\sum_{i=1}^{N} q_i p_i\right)-Q(p_i,\dot p_i)\right)dt\nonumber\\
&=-\int_{t_1}^{t_2}Q(p_i,\dot p_i)\,dt+\left[\sum_{i=1}^{N} q_i p_i\right]_{t_1}^{t_2}
\end{align}

If we compute $\delta S=0$ the conditions $\delta q_i=0$, $\delta p_i=0$ are satisfied by all the allowed trajectories in the extreme points $t_1, t_2$, for this reason all the boundary terms will be removed. In particular, the second contribution in (36) vanishes, and we get the following expression for the variation of the action integral
\begin{equation}
\delta S=\int_{t_1}^{t_2}\delta Q(p_i,\dot p_i)\,dt=0
\end{equation}

This means that
\begin{align}
\delta S&=\sum_{i=1}^{N}\int_{t_1}^{t_2}\left(\frac{\partial Q}{\partial p_i}\delta p_i+\frac{\partial Q}{\partial \dot p_i}\delta \dot p_i \right)\,dt=0
\end{align}
Integrating by parts the last expression

\begin{align}
\delta S&=\left[\sum_{i=1}^{N}\frac{\partial Q}{\partial \dot p_i}\delta p_i\right]_{t_1}^{t_2}+\sum_{i=1}^{N}\int_{t_1}^{t_2}\left(\frac{\partial Q}{\partial p_i}-\frac{d}{dt}\frac{\partial Q}{\partial \dot p_i}\right)\delta p_i \,dt=0
\end{align}
The first is again a boundary term that does not give contribution. Finally, the condition of stationary action $\delta S=0$, implies for each index i, the following relation

\begin{equation}
\frac{\partial Q}{\partial p_i}-\frac{d}{dt}\frac{\partial Q}{\partial \dot p_i}=0
\end{equation}
\\
Therefore, we have been able to derive the Euler-Lagrange equations for the function $Q(p_i,\dot p_i)$ that we previously found in eq.(9), directly from a variational principle.
It is worth noting that the equations that satisfy the functions $\mathcal{L}$ and $Q$ are symmetric exchanging the role of $q_i$ and $p_i$. Thus, the dynamical potential $Q(p_i,\dot p_i)$ can be interpreted as some sort of  ``dual " function associated to the Lagrangian. It plays the same role than the Lagrangian in the momentum space.

What about the other dynamical potential $J(\dot q_i, \dot p_i)$ that was also introduced in the previous sections? Its physical meaning seems more difficult. Nevertheless, we explicitly showed in (13) that employing this potential it is also possible to obtain in a consistent way the conservation of the Hamiltonian .This suggests that this function encodes the same information that any of the other three mechanical potentials, but the dynamical degrees of freedom are organized in a different way.However, this point needs further clarification and will be the subject of future works.

\section{A mathematical correspondence between mechanical and thermodynamic functions}
\label{sec:one-to-one}
The mathematical analogies between classical mechanics and thermodynamics have been studied in several other works \cite{Pe,Ba,Koc,Che,Raj}. In \cite{Pe} for instance, thermodynamics was formulated in a purely symplectic way, a formulation invariant under general canonical transformations, just as classical mechanics is. It is common the study of the thermodynamic structures and functions from a symplectic or ``mechanical'' point of view. However, to study mechanical structures from a point of view parallel to the traditional thermodynamic formalism is less well explored.

Following this latter scheme, it can be demostrated that Hamilton equations can be written as Maxwell relations \cite{Ba}, which is a remarkable result, but we believe that all the richness of  this picture has not yet been completly exploited. According to this ``thermodynamic approach'', we will establish a one-to-one mapping between the different mechanical and thermodynamic Legendre transformations.

With this aim, it is convenient to start the discussion by noting that the internal energy of a thermodynamic system $\mathcal{U}(S,V)$ has a mathematical structure that resembles the Hamiltonian of a mechanical system. Their differentials are equal to
\begin{eqnarray}
d\mathcal{U}=TdS-PdV\nonumber\\
dH=\dot q dp-\dot p dq \
\end{eqnarray}

Indeed, $PdV $ is just an infinitesimal dynamical work done or received (depending on the sign) by the system. There are situations where $\dot p dq$, can also be interpreted as an infinitesimal work.Thus, following this equivalence, it seems natural to establish the one-to-one correspondence 
\begin{eqnarray}
P\rightarrow\ \dot p\nonumber\\
V\rightarrow\ q \nonumber\\
T\rightarrow\dot q\nonumber\\
S\rightarrow p\nonumber\\
\mathcal{U}(S,V)\rightarrow H(p,q)\
\end{eqnarray}

The next step will generate a one-to-one mapping between the thermodynamic and mechanical functions by means of Legendre transforms. For example, following (42), to the thermodynamical transformation $\mathcal{H}(S,P)=\mathcal{U}(S,V)+PV$, will correspond the mechanical $Q(\dot p,p)=H(p,q)+\dot p q$. The $Q(\dot p,p)$ function is well defined in agreement with (5), and has very interesting symmetry properties under different kinds of transformations, as we have studied in detail in the previous section.

In the same way, the so called Gibbs free energy $\mathcal{G}(T,P)=\mathcal{U}(S,V)-TS+PV$, can be put in direct correspondence with $J(\dot p,\dot q)=H(p,q)+\dot p q-\dot q p$. \
This $J(\dot p,\dot q)$ is exactly the same function that was defined in (4), and its dynamical behaviour was also analyzed in detail. Finally, we have the transformation,
\begin{equation}
\mathcal{F}(T,V)=\mathcal{U}(S,V)-TS \,
\end{equation}

This is the Helmholtz free energy, which following our scheme, will map to the mechanical function,
\begin{equation}
\label{eq:lagrangian}
\mathcal{L}^{\prime}(q,\dot q)=H(q,p)-\dot q p \,
\end{equation}
It is straighforward to show that this last function satisfies
\begin{equation}
d\mathcal{L}^{\prime}=-\dot pdq-pd\dot q \,
\end{equation}
and therefore its physical meaning is clear: it represents a redefinition of the Lagrangian, $\mathcal{L}^{\prime}=-\mathcal{L}$.

Therefore, the redefinition only implies a global change of sign of the Lagrangian. When we deal with conservative foces, we will have a change from $\mathcal{L}=T-V$ to $\mathcal{L}^{\prime}=V-T$. It can be shown that this kind of modification does not have physical consequences, due to the fact that the Euler-Lagrange equations are invariant under the transformation $\mathcal{L}^{\prime}\rightarrow -\mathcal{L}$ 
\begin{equation}
 \displaystyle\frac{\partial \mathcal{L}^{\prime}}{\partial q}-\frac{d}{dt}\frac{\partial \mathcal{L}^{\prime}}{\partial \dot q}=0 \,.
\end{equation}

It is easy to see that this identity can also be derived using (45). However, although this change of sign is dynamically irrelevant, it allows us to establish an interesting analogy between thermodynamic and mechanical Legendre transformations. Since this is an exact analogy, the relationship is symmetric. We can take any of the dynamical functions $H(p,q)$, $\mathcal{L}^{\prime}(q,\dot q)$, $Q(\dot p,p)$, $J(\dot p, \dot q)$ and rebuild by correspondence all the thermodynamic potentials.

In order to make the correspondence more clear and graphic, we can use the following Hasse diagrams. In the first Hasse diagram, the ``magnitude" of the potentials decreases as you go down the lines.\\ 
Thus: $\mathcal{H}=\mathcal{F}+PV+TS$, $\mathcal{G}=\mathcal{F}+PV$, and $\mathcal{U}=\mathcal{F}+TS$. Since the mechanical potentials follow the same pattern, they fit perfectly in a similar diagram.In this latter case the ``magnitude" increases as you go down the lines: $Q=\mathcal{L}^{\prime}+\dot p q+\dot q p$, $J=\mathcal{L}^{\prime}+\dot p q$, and $H=\mathcal{L}^{\prime}+\dot q p$

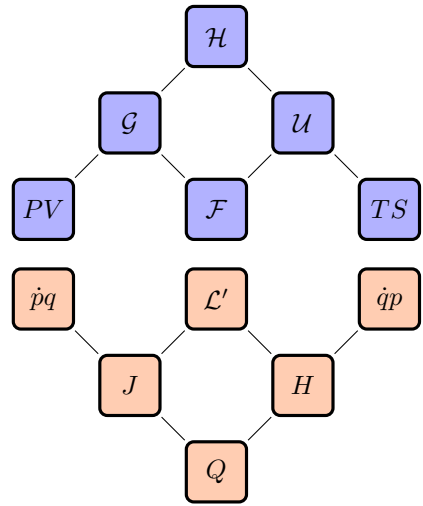
\begin{center}
	\begin{tikzpicture}[scale=1.15]
	\tikzset{packet/.style={rectangle, draw, very thick,
minimum size=8mm, rounded corners=1mm,
fill=blue!30!blue!30}}
 \draw node[packet] (H) at (0,2) {$\mathcal{H}$};
 \draw node[packet] (G) at (-1,1) {$\mathcal{G}$};
 \draw node[packet] (PV) at (-2,0) {$PV$};
 \draw node[packet] (F) at (0,0) {$\mathcal{F}$};
 \draw node[packet] (U) at (1,1) {$\mathcal{U}$};
 \draw node[packet] (TS) at (2,0) {$TS$};
  \draw (H) -- (G);
	\draw(G)-- (PV);
	\draw(G)--(F);
	\draw(H)--(U);
	\draw(F)--(U);
	\draw(U)--(TS);
\end{tikzpicture}
\end{center}

\begin{center}
	\begin{tikzpicture}[scale=1.15]
	\tikzset{packet/.style={rectangle, draw, very thick,
minimum size=8mm, rounded corners=1mm,
fill=red!30!orange!30}}
 \draw node[packet] (H) at (0,2) {$\mathcal{L}^{\prime}$};
 \draw node[packet] (G) at (-1,1) {$J$};
 \draw node[packet] (PV) at (-2,2) {$\dot p q$};
 \draw node[packet] (F) at (0,0) {$Q$};
 \draw node[packet] (U) at (1,1) {$H$};
 \draw node[packet] (TS) at (2,2) {$\dot q p$};
  \draw (H) -- (G);
	\draw(G)-- (PV);
	\draw(G)--(F);
	\draw(H)--(U);
	\draw(F)--(U);
	\draw(U)--(TS);
\end{tikzpicture}
\end{center}

\section{Concluding remarks}
In this work we have analyzed two new Legendre transforms in classical mechanics, the functions $J(\dot q_i,\dot p_i)$, $Q(p_i,\dot p_i)$, which involve the time derivative of the canonical momentum, and we have found that they have very interesting and non-trivial symmetry properties under coordinate transformations. In particular, it is remarkable that one of these functions, the so-called $Q(p_i,\dot p_i)$, can be positioned on equal footing with the Lagrangian of a classical system. On the one hand, it satisfies a particular version of the Euler-Lagrange equations (9). On the other hand, it can be successfully used as well to accomodate Noether's theorem in order to derive general conservation laws. Therefore, we have introduced here an alternative formulation of classical mechanics that complets the usual picture formed by the Lagrangian ($\mathcal{L}$) and the Hamiltonian ($H$), allowing a broader view of the subject. It is always interesting to show new ways of understanding, demonstrating or deriving familiar results in any physical theory. As Feynman said: ``Different equivalent descriptions of the same physics are important because they may lead to different ways to extend them".
With the detailed study of the two new mechanical potentials and their properties we were also able to obtain new insights, like a natural generalization of the Poisson Bracket which was found to be inherent to the formalism introduced.
In addition, we realized that a one-to-one correspondence between the different Legendre transformations of classical mechanics and thermodynamics may be established. This can be achieved paying the price of a dynamically irrelevant (in the case of conservative forces) redefinition of the Lagrangian, from $\mathcal{L}$ to $\mathcal{-L}$.  The important thing is to keep the usual definition of the Hamiltonian as $H=T+V$. But regarding the Lagrangian in the case of conservative forces, taking the usual definition, $\mathcal{L}=T-V$ or the alternative $\mathcal{L}^{\prime}=V-T$, is only a matter of choice that has no physical consequences. The fact that the Lagrangian was originally defined as $\mathcal{L}=T-V$, seems an arbitrary convention taken in absence of more fundamental physical reasons.

\end{document}